\documentclass{ws-mpla}

\begin{document}

\markboth{Forough Nasseri and Sohrab Rahvar} {Dynamics of
Inflationary Cosmology in TVSD Model}

\title{DYNAMICS OF INFLATIONARY COSMOLOGY IN TVSD MODEL}

\author{FOROUGH NASSERI}

\address{Physics Department, Sabzevar University of Tarbiat Moallem,\\
P.O.Box 397, Sabzevar, Iran\\
Khayyam Planetarium, P.O.Box 844, Neishabour, Iran.\\
nasseri@fastmail.fm}

\author{SOHRAB RAHVAR}

\address{Physics Department, Sharif University of Technology,\\
P.O.Box 11365-9161, Tehran, Iran\\
Institute for Studies in Theoretical Physics and Mathematics,\\
P.O.Box 19395-5531, Tehran, Iran\\
rahvar@sina.sharif.ir}

\maketitle

\pub{Received (Day Month Year)}{Revised (Day Month Year)}

\begin{abstract}
Within the framework of a model Universe with time variable space
dimensions (TVSD), known as decrumpling or TVSD model, we study
TVSD chaotic inflation and obtain dynamics of the inflaton, scale factor
and spatial dimension. We also study the quantum fluctuations of the
inflaton field and obtain the spectral index and its running in this
model. Two classes of examples have been studied and comparisons made
with the standard slow-roll formulae. We compare our results with the
recent Wilkinson Microwave Anisotropy Probe (WMAP) data.

\end{abstract}

\section{Introduction}

One of the most intriguing challenges in modern physics is to find
observable consequences of different kinds of theories in higher
dimensions. We here present an inflationary model, known as
decrumpling inflation model in which the number of spatial
dimension has a dynamical behavior and decreases during the expansion
of the Universe with a rate to be less than about
$10^{-14} {\rm yr}^{-1}$.\cite{1}

Our motivation to study decrumpling inflation is
to investigate cosmological implications of time
variability of the number of spatial dimensions.
To do so, we compute the spectral index
and its running within the framework of decrumpling
inflation. For more details about a model Universe with time variable
space dimensions (TVSD), known as TVSD or decrumpling model,
see Refs. [1-6].

To do this research an important conceptual
issue is properly dealt with the meaning of time variability of the
number of spatial dimensions. In Ref. [5], this conceptual issue has
been discussed in detail.

Although time variability of spatial dimensions have not been firmly
achieved in experiments and theories, such dynamical behavior of the
spatial dimensions should not be ruled out in the context of cosmology
and astroparticle physics.

Here, we will be concerened with the approaches proposed in the pioneer
paper\cite{12} where the cosmic expansion of the Universe is named
decrumpling expansion and is due to decrease of the number of spatial
dimensions.

The most important difference between decrumpling model and other
attempts about the time evolution of spatial dimension is that in this
model the number of extra spatial dimensions changes with time while
in other theories the size of extra spatial dimensions is a dynamical
parameter. Based on time variability of the size of
spatial dimensions it has been reported\cite{1} that
the present rate of change of the mean radius of any additional spatial
dimensions to be less than about $10^{-19} {\rm yr^{-1}}$. It is worth
mentioning that this result is based on dynamical behavior
of the size of extra spatial dimensions while in decrumpling model we
take the size of extra spatial dimensions to be constant and the number
of spatial dimensions decreases continuously as the Universe expands.
The present rate of time variation of the number of the spatial dimensions
in decrumpling or TVSD model is about $10^{-13} {\rm yr^{-1}}$.\cite{1}

Another subject which lately has attracted much attention is
the scalar spectral index and its running.

Primordial perturbations from inflation currently provide our only
complete model for the generation of structure in the Universe.
It is commonly stated that a generic prediction of inflationary models
is a scale-invariant spectrum of adiabatic perturbations, characterized
by a scalar spectral index $n_S$ that obeys $n_S-1=0$. However,
this statement is only true for very special spacetimes like a pure
de Sitter spacetime, which does not describe our cosmological history.
For nearly all realistic inflationary models, the value of $n_S$ will
vary with the wave number $k$.

Typically, since $n_S-1\simeq 0$ on the scales probed by the cosmic
microwave background (CMB), the deviations from a constant $n_S$
must be small. Nevertheless, increasingly accurate cosmological
observations provide information about the scalar spectral index on
scales below those accessible to anisotropy measurements of the CMB.
Our such a wide range of scales, it is entirely possible that $n_S$
will exhibit significant running, a value that depends on the scale
on which it is measured. Such running is quantified by the derivative
$dn_S/d\ln k$ and, in fact, recently released data from the
WMAP satellite indicates that\cite{7}
\begin{eqnarray}
\label{1}
n_S\;(k_0=0.05\;\;{\rm Mpc}^{-1})&=&0.93^{+0.03}_{-0.03},\\
\label{2}
\frac{dn_S}{d\ln k}\;(k_0=0.05\;\;{\rm Mpc}^{-1})&=&
-0.031^{+0.016}_{-0.018}.
\end{eqnarray}

The limits on the $n_S$ and its running using WMAP data alone are\cite{8}
\begin{eqnarray}
\label{3}
n_S\;(k_0=0.002\;\;{\rm Mpc}^{-1})&=&1.20^{+0.12}_{-0.11},\\
\label{4}
\frac{dn_S}{d\ln k}\;(k_0=0.002\;\;{\rm Mpc}^{-1})&=&
-0.077^{+0.050}_{-0.052},
\end{eqnarray}
where $k_0$ is some pivot wave number.

Recent data, including that from the WMAP data satellite, show some
evidence that the index runs (changes as a function of the scale $k$
at which it is measured) from $n_S>1$ (blue) on long scales to
$n_S<1$ (red) on short scales. The authors of Ref.[9] investigated
the extent to which inflationary models can accommodate such significant
running of $n_S$. They show that within the slow-roll approximation,
the fact that $n_S-1$ changes sign from blue to red forces the slope
of the potential to reach a minimum at a similar field location.

Chaotic inflation models are usually based upon potentials of the form
$V(\phi)=a \phi^b$ ($a$ is a constant and $b=2,4,...$ is an even integer).
For the interesting cases of $b=2$ and $4$, it has been shown that\cite{10}
\begin{eqnarray}
\label{5}
\frac{dn_S}{d\ln k}=-0.8\times 10^{-3}\;(b=2,\;{\cal{N}}=50),\\
\label{6}
\frac{dn_S}{d\ln k}=-1.2\times 10^{-3}\;(b=4,\;{\cal{N}}=50),
\end{eqnarray}
where ${\cal{N}}$ is the e-folding number.

We will use the natural units system that sets $k_B$, $c$, and $\hbar$
all equal to one, so that $\ell_P=M_P^{-1}=\sqrt{G}$.
To read easily this article we also use the notation $D_t$ instead of
$D(t)$ that means the space dimension $D$ is as a function of time.

The plan of this article is as follows. In section 2, we give a brief
review of decrumpling or TVSD model.
In section 3, we present the dynamical solutions of $\phi(t)$,
$a(t)$ and $D_t(t)$. In section 4, we first present the explicit and
general formulae for the spectral index and its running
within the framework of decrumpling or TVSD inflation and then apply them
to two classes of examples of the inflaton potential.
Finally, we discuss our results and conclude in section 5.

\section{Review of Decrumpling or TVSD Model}

Decrumpling model is based on the assumption that the
basic blocks of the space-time are fractaly
structured.\cite{{5},{12}} In the pioneer paper\cite{12}
the spatial dimension of the Universe was considered as a continuous
time dependent variable. As the Universe expands, its spatial dimension
decreases continuously, thereby generating what has been named a
decrumpling Universe. Then this model has been overlooked and the quantum
cosmological aspects, as well as, a possible test theory for studying
time evolution of Newton's constant have also been
discussed.\cite{{1},{6}} Chaotic inflation in decrumpling model
and its dynamical solutions have also been studied.\cite{{2},{3},{4}}

The concept of decrumpling expansion of the Universe
is inspired by the idea of decrumpling coming from polymer
physics.\cite{{5},{12}} In this model the fundamental building blocks of
the Universe are like cells with arbitrary dimensions having in
each dimension a characteristic size $\delta$ which maybe of the order of
the Planck length ${\mathcal O}$($10^{-33}$ cm) or even smaller
so that the minimum physical radius of the Universe is $\delta$.
These ``space cells'' are embedded in a ${\mathcal D}$ space, where
${\mathcal D}$ may be up to infinity. Therefore, the space dimensions of
the Universe depend on how these fundamental cells are configured in this
embedding space. The Universe may have begun from a very crumpled
state having a very high dimension ${\mathcal{D}}$ and a size $\delta$,
then have lost dimension through a uniform decrumpling which we see
like a uniform expansion. The expansion of space, being now understood like
a decrumpling of cosmic space, reduces the space-time dimension continuously
from ${\mathcal{D}}+1$ to the present value $D_0+1$.
In this picture, the Universe can have any space dimension.
As it expands, the number of spatial dimensions decreases continuously.
The physical process that causes or necessitates such a decrease in
the number of spatial dimensions comes from how these fundamental cells
are embedded in a ${\mathcal{D}}$ space.

As an example, take a limited number of small three-dimensional beads.
Depending on how these beads are embedded in space they can configure to a
one-dimensional string, two-dimensional sheet, or three-dimensional sphere.
This is the picture we are familiar with from the concept of crumpling
in polymer physics where a crumpled polymer has a dimension more
than one. Or take the picture of a clay which can be like a
three-dimensional sphere, or a two-dimensional sheet, or even a
one-dimensional string, a picture based on the theory of fluid membranes.

While it is common to make {\it ad hoc} assumptions in cosmological model
building in the absence of a complete theory of quantum gravity,
some of the particular ingredients of the model owe their physical basis
perhaps more to polymer physics than to cosmology.
Progress with decrumpling model can only be made if there is a breakthrough
in terms of finding a natural mechanism for varying the number of
spatial dimensions in some alternative fashion to that which is
considered here.\cite{1}

\subsection{Relation between the effective space dimension
$D_t(t)$ and characteristic size of the Universe $a(t)$}
\noindent
Assume the Universe consists of a fixed number $\mathbb{N}$ of universal
cells having a characteristic length $\delta$ in each of their
dimensions. The volume of the Universe at the time $t$ depends
on the configuration of the cells. It is easily seen that\cite{5}
\begin{equation}
\label{7}
{\rm vol}_{D_t}({\rm cell})={\rm vol}_{D_0}({\rm cell})\delta^{D_t-D_0},
\end{equation}
where the $t$ subscript in $D_t$ means that $D$ to be as a function
of time, i.e. $D(t)$.

Interpreting the radius of the Universe, $a$, as the radius of
gyration of a crumpled ``universal surface'',
the volume of space can be written\cite{5}
\begin{eqnarray}
\label{8}
a^{D_t}&=&\mathbb{N}\,{\rm vol}_{D_t}({\rm cell})\nonumber\\
   &=&\mathbb{N}\,{\rm vol}_{D_0}({\rm cell}) \delta^{{D_t}-D_0}\nonumber\\
   &=&{a_0}^{D_0} \delta^{{D_t}-D_0}
\end{eqnarray}
or
\begin{equation}
\label{9}
\left( \frac{a}{\delta} \right)^{D_t}=
\left( \frac{a_0}{\delta} \right)^{D_0} = e^C,
\end{equation}
where $C$ is a universal positive constant. Its value has a strong
influence on the dynamics of spacetime, for example on the dimension
of space, say, at the Planck time. Hence, it has physical and cosmological
consequences and may be determined by observation. The zero subscript in
any quantity, e.g. in $a_0$ and $D_0$, denotes its present value.
We coin the above relation as a``dimensional constraint" which relates
the ``scale factor" of decrumpling model to the spatial dimension.
We consider the comoving length of the Hubble
radius at present time to be equal to one. So the interpretation of the
scale factor as a physical length is valid.
The dimensional constraint can be written in this form\cite{5}
\begin{equation}
\label{10}
\frac{1}{D_t}=\frac{1}{C}\ln \left( \frac{a}{a_0} \right) + \frac{1}{D_0}.
\end{equation}
It is seen that by the expansion of the Universe, the space
dimension decreases.
Time derivative of (\ref{9}) or (\ref{10}) leads to
\begin{equation}
\label{11}
{\dot D}_t=-\frac{D_t^2 \dot{a}}{Ca}.
\end{equation}
It can be easily shown that the case of constant space dimension
corresponds to when $C$ tends to infinity. In other words,
$C$ depends on the number of fundamental cells. For $C \to +\infty$,
the number of cells tends to infinity and $\delta\to 0$.
In this limit, the dependence between the space dimensions and
the radius of the Universe is removed, and consequently we
have a constant space dimension.\cite{5}

\subsection{Physical meaning of $D_P$}

We define $D_{P}$ as the space dimension of the Universe when
the scale factor is equal to the Planck length $\ell_{P}$.
Taking $D_0=3$ and the scale of the Universe today to be the present
value of the Hubble radius $H_0^{-1}$ and the space dimension at the
Planck length to be $4, 10,$ or $25$, from Kaluza-Klein and superstring
theory, we can obtain from (\ref{9}) and (\ref{10}) the corresponding
value of $C$ and $\delta$
\begin{eqnarray}
\label{12}
\frac{1}{D_P}&=&\frac{1}{C} \ln \left( \frac{\ell_P}{a_0}
\right) + \frac{1}{D_0}
= \frac{1}{C} \ln \left( \frac{\ell_P}{{H_0}^{-1}} \right) +\frac{1}{3},\\
\label{13}
\delta &=& a_0 e^{-C/D_0}= H_0^{-1} e^{-C/3}.
\end{eqnarray}
In Table 1, values of $C$, $\delta$ and also ${\dot D}_t|_0$ for some
interesting values of $D_P$ are given.
These values are calculated by assuming $D_0=3$ and
$H_0^{-1}=3000 h_0^{-1} {\rm Mpc}=9.2503 \times 10^{27} h_0^{-1}{\rm cm}$,
where we take $h_0=1$.\cite{5}

\section{Dynamical Solutions of $\phi(t)$, $a(t)$, and $D_t(t)$}

Using the slow-roll approximation, equations of motion of $a(t)$,
$D_t(t)$, and $\phi(t)$ for the potential
$V(\phi)=m^2\phi^2/2$ are given by\cite{3}
\begin{eqnarray}
\label{14}
&&\left( \frac{\dot a}{a} \right)^2 \simeq \frac{8 \pi m^2 \phi^2}
{D_t(D_t-1)M_P^2},\\
\label{15}
&& \frac{D_t^2 \dot{a} \dot{\phi}}{a}\left[ \frac{1}{D_0} -
\frac{1}{C} \left\{ \ln \chi_c + \frac{1}{2} \ln \pi - \frac{1}{2}
\psi \left( \frac{D_t}{2} +1 \right) \right\} \right] \simeq -m^2 \phi,\\
\label{16}
&&{{\dot D}_t}^2 \simeq \frac{8 \pi D_t^3 m^2 \phi^2}{C^2 (D_t-1) M_P^2},
\end{eqnarray}
where $m$ is the inflaton mass, $m=1.21 \times 10^{-6} M_P$.
In natural unit system $M_P=2.18 \times 10^{-5} {\rm gr}$
and $1 {\rm gr}=8.52 \times 10^{47} \sec^{-1}$.
So the Planck mass is $M_P=1.85 \times 10^{43} \sec^{-1}$
and the inflaton mass is $m=2.24 \times 10^{37} \sec^{-1}$.\cite{13}

In [3] we obtained dynamical solutions of Eqs.({\ref{14}-{16}}).
Using definition of the following parameters\cite{3}
\begin{eqnarray}
\label{17}
\alpha &\equiv& -\frac{m D_0}{2 D_i} \sqrt{\frac{D_i-1}{2 \pi D_i}},\\
\label{18}
\beta &\equiv& \frac{m(2D_i-3)}{(D_i-1)}\sqrt{\frac{\pi D_i}{2(D_i-1)}},
\end{eqnarray}
and
\begin{equation}
\label{19}
\gamma \equiv -\alpha + \frac{\beta}{C} \left( \frac{\phi_i}{M_P}
\right)^2,
\end{equation}
the solutions of Eqs.({\ref{14}-{16}}) are given by\cite{3}
\begin{eqnarray}
\label{20}
\frac{\phi(t)}{M_P}&=&\sqrt{\frac{\gamma C}{\beta}}
\tanh \left[ - \sqrt{\frac{\beta \gamma}{C}} (t-t_i)
+{\rm arctanh} \left( \frac{\phi_i}{M_P}
\sqrt{\frac{\beta}{C \gamma}} \right) \right]+...,\\
\label{21}
D_t(t) & = & \Bigg( 1 + \Bigg\{ 1 - \frac{2}{D_i}
\left( 1 - \frac{1}{2D_i} \right)
- \frac{8 \pi}{C D_0} \Bigg[ \left( \frac{\phi_i}{M_P} \right)^2\nonumber\\
&-&\frac{\gamma C}{\beta} \tanh^2 \left[ - \sqrt{\frac{\beta \gamma}{C}}
(t-t_i) + {\rm arctanh} \left( \frac{\phi_i}{M_P}
\sqrt{\frac{\beta}{C \gamma}} \right) \right] \Bigg] \Bigg\}^{1/2}
\Bigg) \nonumber\\
&\times& \Bigg\{ \frac{2}{D_i} \left( 1- \frac{1}{2D_i} \right)
+ \frac{8 \pi}{C D_0} \Bigg[ \left( \frac{\phi_i}{M_P} \right)^2
-\frac{\gamma C}{\beta} \tanh^2 \Bigg[ - \sqrt{\frac{\beta \gamma}{C}}
(t-t_i)\nonumber\\
&+&{\rm arctanh} \left( \frac{\phi_i}{M_P}
\sqrt{\frac{\beta}{C \gamma}} \right) \Bigg] \Bigg] \Bigg\}^{-1},\\
\label{22}
a(t) &=& a_i \exp \left(\frac{8 \pi D_t [\phi_i^2-\phi^2(t) ]}
{D_0 M_P^2 [ (2-\frac{1}{D_i}) D_t -1]} \right).
\end{eqnarray}
The $i$ subscript in $D_i$, $\phi_i$ and $a_i$ are their values
at the beginning of inflationary epoch. Their values are given in
Ref. [3], see Table 1. Inserting the values of $D_0=3$, $\phi_i$
and $D_i$ in equations (\ref{17}), (\ref{18}) and (\ref{19}), one can
easily obtain the values of $\alpha$, $\beta$ and $\gamma$, see Table 1.
\begin{table}
\tbl{Values of $D_P$, $C$, $\delta$, ${\dot D}_t|_0$, $D_i$, $\phi_i/M_P$,
$\alpha$, $\beta$, and $\gamma$. $D_P=3,\,4$ and $10$ are corresponding to
$a_i=1.02 \times 10^{-25},\,1.06 \times 10^{-30}$, and $2.28
\times 10^{-43} {\rm cm}$, respectively.}
{\begin{tabular}{@{}ccccccccc@{}} \toprule
$D_P$ & $C$ & $\delta$(cm) & ${\dot D}_t|_0({\rm yr}^{-1})$ &
$D_i$ & $\phi_i/M_P$ & $\alpha$ & $\beta$ & $\gamma$ \\ \hline
\hline
$3$ &  $+\infty$ & $0$ & $0$ & $3.00$ & $3.060$ & $-0.16m$ & $2.30m$ & $0.16m$\\
$4$ &  $1678.8$ & $8.6 \times 10^{-216}$ & $-5.48 \times 10^{-13}$ & $3.94$ & $3.491$ & $-0.13m$ & $2.41 m$ & $0.15m$\\
$10$ & $599.57$ & $1.5 \times 10^{-59}$ & $-1.53 \times 10^{-12}$ & $16.08$ & $4.512$ & $-0.04m$ & $2.50m$ & $0.12m$\\
\hline
\end{tabular}}
\end{table}
When the space dimension at the Planck length to be equal to
$D_P=4$ corresponding to $C=1678.8$, one can easily obtain
the dynamical solutions of $\phi(t)$, $D_t(t)$ and $a(t)$, which are given by
\begin{eqnarray}
\label{23}
\frac{\phi(t)}{M_P} &=& 10.22 \tanh \left[ -2.24 \times 10^{35}(t-t_i)
+{\rm arctanh\,0.34} \right],\\
\label{24}
D_t(t) &=& \Bigg( 1+ \Bigg\{ 0.56 - 4.99 \times 10^{-3} \Bigg[
12.19 -104.49 \tanh^2 \Bigg[ -2.24 \times 10^{35} (t-t_i)\nonumber\\
&+&{\rm arctanh}\,0.34 \Bigg] \Bigg] \Bigg\} ^{1/2} \Bigg)
\times \Bigg\{ 0.44 + 4.99 \times 10^{-3}  \Bigg[12.19 -104.49 \nonumber\\
&\times&\tanh^2 \Bigg[ -2.24 \times 10^{35} (t-t_i)
+{\rm arctanh}\,0.34 \Bigg] \Bigg] \Bigg\}^{-1},\\
\label{25}
a(t) &=& 1.06 \times 10^{-30} \exp \left[ \frac{8 \pi D_t
\left(12.19 - (\frac{\phi(t)}{M_P})^2 \right)}
{3 (1.75D_t-1)} \right].
\end{eqnarray}
Using Equations (\ref{17})-(\ref{22}), for the space dimension at the
Planck length $D_P$ to be equal to $10$ corresponding to $C=599.57$, one
can easily obtain the dynamical solutions of $\phi(t)$, $D_t(t)$ and $a(t)$,
which are given by
\begin{eqnarray}
\label{26}
\frac{\phi(t)}{M_P}&=&5.36 \tanh \left[ -4.48 \times 10^{35}
(t-t_i) + {\rm arctanh}\,0.84 \right],\\
\label{27}
D_t(t)&=& \Bigg( 1 + \Bigg\{ 0.88 - 0.01 \Bigg[ 20.36 - 28.78
\times \tanh^2 \Bigg[ -1.12 \times 10^{34} (t-t_i) \nonumber\\
&+& {\rm arctanh}\,0.84\Bigg] \Bigg] \Bigg\}^{1/2} \Bigg)
\times \Bigg\{ 0.12 + 0.01 \Bigg[ 20.36 - 28.78 \nonumber\\
&\times&\tanh^2\Bigg[ - 4.48 \times 10^{35} (t-t_i)
+ {\rm arctanh}\,0.84 \Bigg] \Bigg]\Bigg\}^{-1},\\
\label{28}
a(t)&=&2.28 \times 10^{-43} \exp \left( \frac{8 \pi D_t
[20.36 - (\frac{\phi(t)}{M_P})^2 ]}{3[1.94D_t-1]} \right).
\end{eqnarray}
\begin{figure}[th]
\centerline{\psfig{file=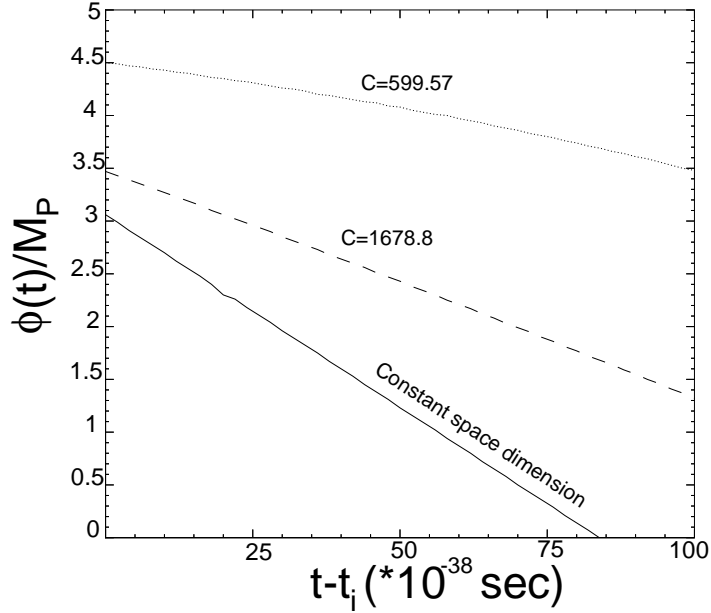,width=10.5cm}}
\vspace*{8pt}
\caption{Inflaton field as a function of time for $C=1678.8$ (dashed line),
$C=599.57$ (dotted line), and three constant space dimension or
$C \to +\infty$ (solid line).}
\end{figure}

\begin{figure}[th]
\centerline{\psfig{file=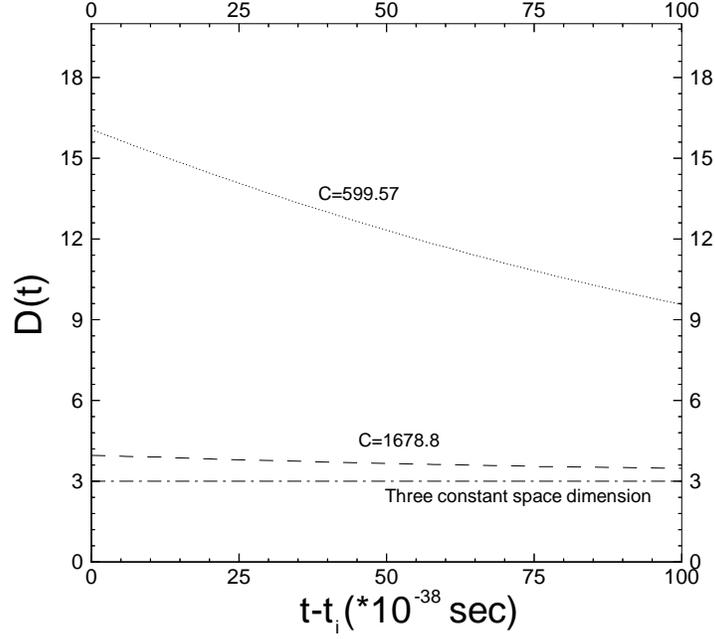,width=10.5cm}}
\vspace*{8pt}
\caption{Space dimensions as a function of time for $C=1678.8$ (dashed line),
$C=599.57$ (dotted line), and three constant space dimension or
$C \to +\infty$ (dashdot line).}
\end{figure}

\begin{figure}[th]
\centerline{\psfig{file=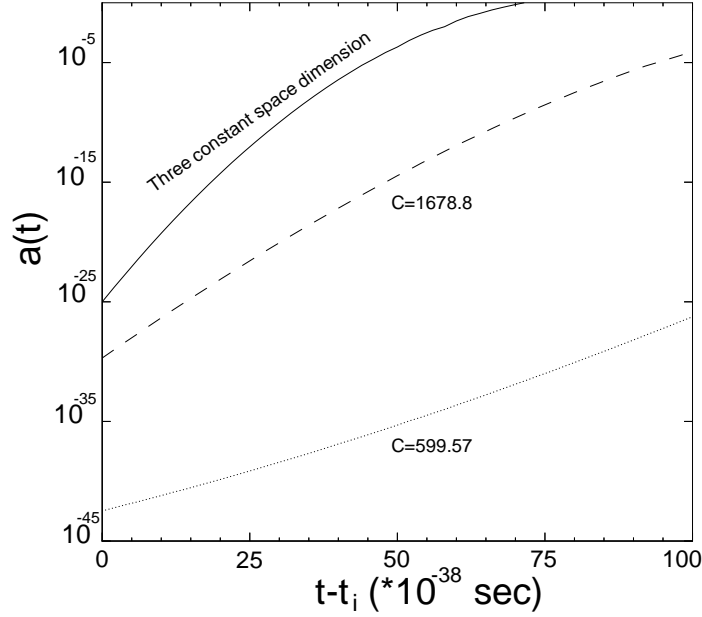,width=10.5cm}}
\vspace*{8pt}
\caption{Scale factor as a function of time for $C=1678.8$ (dashed line),
$C=599.57$ (dotted line), and three-constant space dimension or
$C \to +\infty$ (solid line). We use the log scale for the axis of $a(t)$.
Here the scale factor is the same as the physical length because we take
the comoving length of the Hubble radius to be equal to one.}
\end{figure}

In three constant space dimension corresponding to $C \to +\infty$,
one can use Eqs. (\ref{17})-(\ref{22}) and obtain the following dynamical
solutions of $\phi(t)$ and $a(t)$ which are given by
\begin{eqnarray}
\label{29}
\frac{\phi(t)}{M_P}&=& 3.060 - 3.65 \times 10^{36}(t-t_i),\\
\label{30}
a(t)&=& 1.02 \times 10^{-25}\exp \left( {2 \pi}
\left[ 9.36- \left( \frac{\phi(t)}{M_P} \right)^2 \right] \right).
\end{eqnarray}
Using Eqs. (\ref{23}), (\ref{26}), and (\ref{29}), we plot the time
evolution of the inflaton field in Fig.(1). In Fig.(2), time evolution
of the spatial dimension has been shown by Eqs. (\ref{24}) and (\ref{27}).
Finally in Fig.(3), the time evolution of the scale factor has
been shown by Eqs. (\ref{25}), (\ref{28}) and (\ref{30}). It is worth
mentioning that we use log scale for the axis of the scale factor in
Fig.(3). As mentioned in [3], since we take the comoving length of
the Hubble radius to be equal to one, so the scale factor is equal to
the physical length. As shown in [3], the inflationary epoch
lasts about $1.003 \times 10^{-36} \sec$ and $1.159 \times
10^{-36} \sec$ for $D_P=4$ and $D_P=10$, respectively.
In three constant space dimension the
inflationary epoch lasts about $7.598 \times 10^{-37} \sec$.\cite{3} We
have used these values of time for plotting Figs.(1), (2) and (3).

\section{Running of the Spectral Index in Decrumpling or TVSD Inflation}

In this section, we present explicit and general formulae for the
spectral index and its running within the framework of TVSD or
decrumpling inflation.
For the purposes of illustration, we apply our results for
two classes of examples of the inflaton potential.

\subsection{Decrumpling or TVSD chaotic inflation}

Inflation has been studied in the framework of decrumpling or TVSD
model. The crucial equations are\cite{{3},{4}}
\begin{eqnarray}
\label{31}
&&H^2 = \left( \frac{\dot a}{a} \right)^2 =
\frac{16 \pi}{D_t(D_t-1)M_P^2} \left( \frac{1}{2} {\dot\phi}^2 +
V(\phi) \right) -\frac{k}{a^2},{\mbox{Friedmann equation}},\\
\label{32}
&&\ddot{\phi}+D_tH\dot\phi+{\dot D}_t \dot\phi \left( \ln \frac{a}{a_0} +
\frac{d \ln V_{D_t}}{dD_t} \right) = -V'(\phi),\;\;\;\;\;
{\mbox{Fluid equation}},
\end{eqnarray}
where $V_{D_t}$ is the volume of the space-like sections\cite{5}
\begin{eqnarray}
\label{33} V_{D_t} &=&
\begin{cases}
\frac{2 \pi^{(D_t+1)/2}}{\Gamma[(D_t+1)/2]},     & if
$\;k=+1,\;\;\;{\mbox{closed decrumpling model,}}$ \cr
 \frac{\pi^{(D_t/2)}}{\Gamma(D_t/2+1)}{\chi_c}^{D_t},  & if $\;k=0,\;\;\;{\mbox{flat decrumpling model,}}$ \cr
               \frac{2\pi^{(D_t/2)}}{\Gamma(D_t/2)}f(\chi_c),  & if $\;k=-1,\;\;\;{\mbox{open decrumpling model.}}$
               \cr
\end{cases}
\end{eqnarray}
These volumes of space-like sections are valid even in the case
of constant $D$-space.\cite{5}
Here $\chi_C$ is a cut-off and $f(\chi_c)$ is a function
thereof.\cite{2}

Using the slow-roll approximation in decrumpling model\cite{{3},{4}}
\begin{eqnarray}
\label{34}
{\dot{\phi}}^2 &\ll& V(\phi),\\
\label{35}
{\ddot{\phi}} &\ll& D_tH{\dot \phi},\\
\label{36}
-{\dot H} &\ll& H^2
\end{eqnarray}
and
\begin{equation}
\label{37}
{\dot D}_t \left( \ln \frac{a}{a_0} +
\frac{d \ln V_{D_t}}{d D_t} \right) \ll D_t H,
\end{equation}
Eqs. (\ref{31}) and (\ref{32}) can be rewritten for a flat decrumpling
model, i.e. $k=0$, in the simpler set
\begin{eqnarray}
\label{38}
H^2 &\simeq& \frac{16 \pi V(\phi)}{D_t(D_t-1) M_P^2},\\
\label{39}
D_t H \dot{\phi} &\simeq& - V'(\phi).
\end{eqnarray}
Note that the slow-roll condition (\ref{37}) has not been considered
in Refs. [3,4]. The validity of this
condition is obvious by regarding Eq. (\ref{11}). Substituting (\ref{11})
in (\ref{37}), dynamics of the spatial dimension is given by\cite{3,4}
\begin{equation}
\label{40}
{{\dot D}_t}^2 \simeq  \frac{16 \pi D_t^3 V(\phi)}{C^2 (D_t -1) M_P^2}.
\end{equation}
During inflation, $H$ is slowly varying in the sence that its change per
Hubble time $\epsilon \equiv -{\dot H}/{H^2}$ is less than one.
The slow-roll condition $|\eta| \ll 1$ is actually a consequences of the
condition $\epsilon \ll 1$ plus the slow-roll approximation
$D_t H {\dot \phi} \simeq - V'(\phi)$. Deferentiating (\ref{39}) one finds
\begin{equation}
\label{41}
\frac{\ddot \phi}{H \dot{\phi}} = \epsilon - \eta + \frac{D_t}{C},
\end{equation}
where the slow-roll parameters in decrumpling model are defined by
\begin{eqnarray}
\label{42}
\epsilon &\equiv& \frac{(D_t-1) M_P^2}{32 \pi}
\left( \frac{V'}{V} \right)^2,\\
\label{43}
\eta &\equiv& \frac{(D_t-1) M_P^2}{16 \pi} \left( \frac{V''}{V} \right).
\end{eqnarray}
It should be emphasized that the slow-roll parameters in decrumpling model
as presented in Refs. [3,4] are different from those given in (\ref{42})
and (\ref{43}).
This difference is due to the slow-roll condition (\ref{37}) which has
not been considered in Refs. [3,4]. Furthermore, in the constant
$D$-space, the slow-roll parameters (\ref{42}) and (\ref{43}) are also
valid by substituting $D_t$ by $D$, see Refs. [3,4].

\subsection{Explicit formulae for running in decrumpling or TVSD
inflation}

The amplitudes of scalar and tensor perturbations generated in
decrumpling or TVSD inflation can be expressed by\cite{11,14}
\begin{equation}
\label{44}
A_S^2=\left(\frac{H}{2\pi}\right)^2
\left( \frac{H}{\dot{\phi}} \right)^2,
\end{equation}
and
\begin{equation}
\label{45}
A_T^2=\frac{8 \pi}{M_P^2 (D_t-1)} \left( \frac{H}{2 \pi} \right)^2.
\end{equation}
These amplitudes are equal to $\frac{25}{4}$ times the
amplitudes as given in Ref. [14].
The amplitudes of scalar and tensor perturbations generated in
decrumpling inflation can be determined by substituting (\ref{38})
and (\ref{39}) in (\ref{44}) and (\ref{45})
\begin{eqnarray}
\label{46}
A_S^2 &=& \frac{9216 \pi V^3}{D_t^3 (D_t-1)^3 M_P^6 V'^2},\\
\label{47}
A_T^2 &=& \frac{32 V}{D_t (D_t -1)^2 M_P^4}.
\end{eqnarray}
These expressions are evaluated at the horizon crossing time when
$k=aH$. Since the value of Hubble constant does not change too much during
inflationary epoch, we can obtain $dk = Hda$ and
$d\ln k = Hdt = da/a$ . Using the slow-roll
condition in decrumpling inflation
\begin{equation}
\label{48}
\frac{d}{d \ln k} = - \frac{V'}{D_t H^2} \frac{d}{d \phi},
\end{equation}
and also the dimensional constraint (\ref{10}) of the model we have
\begin{eqnarray}
\label{49}
\frac{dD_t}{da}=-\frac{D_t^2}{Ca},\\
\label{50}
\frac{dD_t}{d \ln k} = - \frac{D_t^2}{C},\\
\label{51}
\frac{dD_t}{d\phi}= \frac{D_t^3 H^2}{V' C}.
\end{eqnarray}
After a lengthy but straightforward calculation
by using (\ref{38}), (\ref{39}), (\ref{42}),
(\ref{43}) and (\ref{46})-(\ref{51}) we find
\begin{eqnarray}
\label{52}
n_S-1 &\equiv& \frac{d \ln A_S^2}{d \ln k}
=-6 \epsilon +2 \eta +
\frac{3D_t(2D_t-1)}{C(D_t-1)},\\
\label{53}
n_T &\equiv& \frac{d \ln A_T^2}{d \ln k}
= -2 \epsilon + \frac{D_t(3D_t-1)}{(D_t-1)C}.
\end{eqnarray}
where $n_S$ and $n_T$ are the spectral indices of scalar and tensor
perturbations, respectively. If $n_S$ and $n_T$ are expressed as a
function of e-folding ${\cal{N}}$, one can use the fact that
$\frac{d}{d \ln k}=-\frac{d}{d {\cal{N}}}$ to obtain the desired
derivatives even more easily.

To calculate the running of the scalar spectral index,
we use the following expressions
\begin{eqnarray}
\label{54}
\frac{d \epsilon}{d \ln k} &=& - \frac{D_t^2}{C(D_t-1)}\epsilon -
2 \epsilon \eta + 4 \epsilon^2,\\
\label{55}
\frac{d \eta}{d \ln k} &=& -\frac{D_t^2}{C(D_t-1)} \eta +
2 \epsilon \eta- \xi,
\end{eqnarray}
where the third slow-roll parameter is defined by
\begin{equation}
\label{56}
\xi \equiv \frac{(D_t-1)^2M_P^4}{(16 \pi)^2}\left(\frac{V'V'''}{V^2}\right).
\end{equation}
Using above equations, running of the scalar spectral index
in decrumpling or TVSD inflation has this explicit expression
\begin{equation}
\label{57}
\frac{d n_S}{d \ln k}= 16 \epsilon \eta
- 24 \epsilon^2 - 2 \xi
+\frac{6 D_t^2}{C(D_t-1)} \epsilon-\frac{2D_t^2}{C(D_t-1)} \eta
-\frac{3D_t^2(2D_t^2-4D_t+1)}{C^2(D_t-1)^2}.
\end{equation}
The standard consistency relation between the tensor to scalar
perturbations and the tensor spectral index is
\begin{equation}
\label{58}
R=-8n_T,
\end{equation}
where $R$ is the ratio of tensor to scalar perturbations\cite{14}
\begin{equation}
\label{59}
R \equiv 16 \frac{A_T^2}{A_S^2}.
\end{equation}
This equation can be rewritten in decrumpling or TVSD inflation as
\begin{equation}
\label{60}
R \equiv 16 \frac{A_T^2}{A_S^2} + f \left( \frac{1}{C} \right),
\end{equation}
where $f \left( \frac{1}{C} \right)$ is a function of the inverse of
universal constant of decrumpling or TVSD model, see Eq.(\ref{10}).
From Eqs.(\ref{46}), (\ref{47}) and ({\ref{60}), one can obtain
\begin{equation}
\label{61}
R \equiv \frac{16 D_t^2}{9} \epsilon + f \left( \frac{1}{C} \right).
\end{equation}
The consistency equation in decrumpling or TVSD inflation is
\begin{equation}
\label{62}
R = - \frac{8 D_t^2}{9} n_T.
\end{equation}
In the limit $D_t=3$, Eq.(\ref{62}) yields Eq.(\ref{58}).
From Eqs.(\ref{53}) and (\ref{62}), we have
\begin{equation}
\label{63}
R = \frac{16 D_t^2}{9} \epsilon
- \frac{8 D_t^3 (3D_t-1)}{9(D_t-1)C}.
\end{equation}
Comparing this equation with (\ref{61}) gives
\begin{equation}
\label{64}
f \left( \frac{1}{C} \right)=
- \frac{8 D_t^3 (3D_t-1)}{9(D_t-1)C}.
\end{equation}
Let us now study two inflationary potentials.
We first study $V(\phi)=m^2\phi^2/2$ and then
$V(\phi)=\lambda \phi^4$.

\subsubsection{The first example: $V(\phi)=\frac{1}{2}m^2\phi^2$}

For the purposes of illustration, we now consider the potential
$V(\phi)=m^2 \phi^2/2$

\begin{eqnarray}
\label{65}
\epsilon&=&\eta=\frac{(D_t-1)M_P^2}{8 \pi \phi^2},\\
\label{66}
\xi&=&0.
\end{eqnarray}
Using the definition of e-folding in decrumpling inflation\cite{11}
\begin{equation}
\label{67}
{\mathcal{N}}=-\frac{16\pi}{M_P^2} \int_{\phi}^{\phi_f}
\frac{V}{(D_t-1)V'}d\phi,
\end{equation}
we have
\begin{equation}
\label{68}
{\mathcal{N}}=\frac{4\pi}{(D_t-1)M_P^2}(\phi^2-\phi_f^2),
\end{equation}
where $\phi_f$ is the value of the inflaton field at the end of
inflation. Note that in the integral of (\ref{67}),
$D_t$ is as a function of the inflaton field. To integrate in (\ref{67})
we take $D_t$ to be independent on the inflaton field because
the relationship between $D_t$ and $\phi$ in decrumpling inflation is
too weak.\cite{3} For this reason, our approximation is appropriate for
integration of Eq. (\ref{67}).

To obtain e-folding as a function of the inflaton field, we must obtain
$\phi_f$. From $\epsilon=\eta=1$ we get
\begin{equation}
\label{69}
\phi_f=\sqrt{\frac{D_t-1}{8\pi}}M_P.
\end{equation}
From (\ref{68}) and (\ref{69}), we have
\begin{equation}
\label{70}
\phi^2=\frac{(D_t-1)M_P^2}{8\pi}(2{\mathcal{N}}+1).
\end{equation}
So we are led to the spectral index and its running
\begin{eqnarray}
\label{71}
n_S-1&=&-\frac{4}{(2{\mathcal{N}}+1)}+
\frac{3D_t(2D_t-1)}{C(D_t-1)},\\
\label{72}
\frac{d n_S}{d \ln k}&=& -\frac{8}{(2{\mathcal{N}}+1)^2}-
\frac{4D_t^2}{C(D_t-1)(2{\mathcal{N}}+1)}-
\frac{3D_t^2(2D_t^2-4D_t+1)}{C^2(D_t-1)^2}.
\end{eqnarray}
One can obtain the ratio of tensor to scalar perturbations
by (\ref{63}) and (\ref{65})
\begin{equation}
\label{73}
R = \frac{2D_t^2(D_t-1) M_P^2}{9 \pi \phi^2}
- \frac{8 D_t^3 (3D_t-1)}{9(D_t-1)C}.
\end{equation}
Substituting (\ref{70}) in (\ref{73}) yields
\begin{equation}
\label{74}
R = \frac{16 D_t^2}{9 (2 {\cal{N}} + 1)}
-\frac{8 D_t^3 (3D_t-1)}{9(D_t-1)C}.
\end{equation}
From (\ref{71}) and (\ref{74}), one can obtain
\begin{equation}
\label{75}
n_S-1 = - \frac{9 R}{4 D_t^2}-\frac{D_t}{C(D_t-1)}.
\end{equation}

\subsubsection{The second example: $V(\phi)=\lambda \phi^4$}

For the second example, we study $V(\phi)=\lambda \phi^4$.
The slow-roll parameters are
\begin{eqnarray}
\label{76}
\epsilon &=& \frac{(D_t-1)M_P^2}{2\pi \phi^2},\\
\label{77}
\eta&=&\frac{3(D_t-1)M_P^2}{4\pi\phi^2},\\
\label{78}
\xi&=&\frac{3(D_t-1)^2M_P^4}{8\pi^2\phi^4}.
\end{eqnarray}
Using the definition of e-folding in inflation, we have
\begin{equation}
\label{79}
{\mathcal{N}}=\frac{2\pi}{M_P^2(D_t-1)}\left(\phi^2-\phi_f^2 \right).
\end{equation}
To obtain the e-folding number as a function of the inflaton field,
we must obtain the value of the inflaton field at the end of inflation.
From $\epsilon=1$ we get
\begin{equation}
\label{80}
\phi_f=\sqrt{\frac{D_t-1}{2\pi}}M_P.
\end{equation}
Assuming $\eta=1$ and $\xi=1$,
we obtain
\begin{equation}
\label{81}
\phi_f=\sqrt{\frac{3(D_t-1)}{4\pi}}M_P
\end{equation}
and
\begin{equation}
\label{82}
\phi_f=\left( \frac{3(D_t-1)^2}{8\pi^2} \right)^{1/4}M_P,
\end{equation}
respectively. These values of $\phi_f$ based on the condition
$\eta=1$ and $\xi=1$ are larger than $\phi_f$ arisen from $\epsilon=1$.
We here take the condition $\epsilon=1$ by itself is a true condition to
obtain $\phi_f$. From (\ref{79}) and (\ref{80}) we have
\begin{equation}
\label{83}
\phi^2=\frac{(D_t-1)({\mathcal{N}}+1)M_P^2}{2\pi}.
\end{equation}
So we are led to the spectral index and its running
\begin{eqnarray}
\label{84}
&&n_S-1 = - \frac{3}{({\mathcal{N}}+1)} +\frac{3D_t(2D_t-1)}{C(D_t-1)},\\
\label{85}
&&\frac{dn_S}{d\ln k}=-\frac{3}{({\mathcal{N}}+1)^2}+
\frac{3D_t^2}{C(D_t-1)({\mathcal{N}}+1)}
-\frac{3D_t^2 (2D_t^2-4D_t+1)}{C^2(D_t-1)^2}.
\end{eqnarray}
One can obtain the ratio of tensor to scalar perturbations
by (\ref{63}) and (\ref{76})
\begin{equation}
\label{86}
R=\frac{8 D_t^2 (D_t-1) M_P^2}{9 \pi \phi^2}
-\frac{8 D_t^3(3D_t-1)}{9 (D_t -1)C}.
\end{equation}
Substituting (\ref{83}) in (\ref{86}) yields
\begin{equation}
\label{87}
R=\frac{16 D_t^2}{9 ( {\cal{N}} +1)}
- \frac{8 D_t^3 (3D_t-1)}{9 (D_t -1) C}.
\end{equation}
From (\ref{84}) and (\ref{87}), one can obtain
\begin{equation}
\label{88}
n_S-1 = - \frac{27 R}{16 D_t^2} + \frac{3D_t}{2C}.
\end{equation}

\subsection{Numerical calculations}

The WMAP team\cite{8} carried out the likelihood analysis by varying the
four quantities $A_S$, $R$, $n_S$ and $dn_S/d \ln k$. The quantities
$n_T$ and $dn_T/d \ln k$ are related to those by consistency equation,
and $d n_T/d \ln k$ has anyway always been ignored so far in parameter
fits as its cosmological consequences are too subtle for current
or near-future data to detect. Thus, we ignore it and use
four observables $(A, R, n_S, dn_S/d \ln k)$ as free parameters.

As shown in (\ref{46}), the value of $A_S$ in decrumpling
model is independent of the inverse of universal constant,
$\frac{1}{C}$, of the model. This means that the value of $A_S$
in the standard inflation in $D$-constant spatial dimension
is equal to the value of $A_S$ in decrumpling model.
We therefore conclude that the
characteristic of time variabilty of the space dimension in decrumpling
inflation does not change the value of $A_S$.

\subsubsection{The first potential: $V(\phi)=m^2\phi^2/2$}
Let us calculate the value of $n_S-1$ and $dn_S/d\ln k$
for the potential $m^2\phi^2/2$. Substituting $D_t=3$ for the
present value of the space dimension in Eqs. (\ref{71}) and (\ref{72})
\begin{eqnarray}
\label{89}
n_S-1&=&-\frac{4}{(2{\cal{N}}+1)}+\frac{45}{2C},\\
\label{90}
\frac{dn_S}{d\ln k}&=&-\frac{8}{(2{\cal{N}}+1)^2}-
\frac{18}{C(2{\cal{N}}+1)}-\frac{189}{4C^2}.
\end{eqnarray}
From Eqs. (\ref{89}) and (\ref{90}),
one can obtain for $C=599.57$
corresponding to $D_P=10$, varying $(n_S-1)$
from $-0.00207$ to $0.00916$ and ${\cal{N}}$ from $50$ to $70$,
$d n_S/d \ln k$ varies from about $-0.00121$ to
almost $-0.00075$. In this case, the value of $n_S-1$ vanishes
within about ${\cal{N}}=52.8$ e-folding. In other words, the sign of
$n_S-1$ within about ${\cal{N}}=52.8$ changes from red ($n_S<1$)
to blue ($n_S>1$) tilt.

We obtain for $C=1678.8$ corresponding to $D_P=4$, varying $(n_S-1)$
from $-0.0262$ to $-0.0150$ and ${\cal{N}}$ from $50$ to $70$,
$d n_S/d \ln k$ varies from about $-0.0009$ to
almost $-0.0005$.

For $C \to +\infty$ corresponding to the constant space dimension
and $D_P=3$, varying $(n_S-1)$
from $-0.03960$ to $-0.02836$ and ${\cal{N}}$ from $50$ to $70$,
$d n_S/d \ln k$ varies from about $-0.00078$ to
almost $-0.00040$.

It is important to study the ratio of tensor to scalar
perturbations as a function of $n_S$ to check whether or not this
ratio lies within the observational sigma bounds.\cite{14}
From (\ref{75}), one can obtain in three-space dimension
\begin{equation}
\label{91}
n_S-1 = -\frac{R}{4}-\frac{3}{2C}.
\end{equation}
To plot $n_S$ and $R$ together with the $2D$ posterior constraints
in the $n_S-R$ plane, we need to run the CAMB program coupled to
the CosmoMc (Cosmological Monte Carlo) code, for more details see
Ref.[14] and references therein. We will study this problem in
our future works.

\subsubsection{The second potential: $V(\phi)=\lambda \phi^4$}
We now calculate the value of $n_S-1$ and $dn_S/d\ln k$
for the potential $\lambda \phi^4$. Substituting $D_t=3$ for the
present value of the space dimension in Eqs. (\ref{84}) and (\ref{85})
\begin{eqnarray}
\label{92}
n_S-1&=&-\frac{3}{({\cal{N}}+1)}+\frac{45}{2C},\\
\label{93}
\frac{dn_S}{d\ln k}&=&-\frac{3}{({\cal{N}}+1)^2}
+\frac{27}{2C({\cal{N}}+1)}-\frac{189}{4C^2}.
\end{eqnarray}

From Eqs. (\ref{92}) and (\ref{93}), one can obtain
for $C=599.57$ corresponding to $D_P=10$, varying $(n_S-1)$
from $-0.02130$ to $-0.00473$ and ${\cal{N}}$ from $50$ to $70$,
$d n_S/d \ln k$ varies from about $-0.00084$ to
almost $-0.00041$.

We obtain for $C=1678.8$ corresponding to $D_P=4$, varying $(n_S-1)$
from $-0.04542$ to $-0.02885$ and ${\cal{N}}$ from $50$ to $70$,
$d n_S/d \ln k$ varies from about $-0.00101$ to
almost $-0.00050$.

For $C \to +\infty$ corresponding to the constant space dimension
and $D_P=3$, varying $(n_S-1)$
from $-0.05882$ to $-0.04225$ and ${\cal{N}}$ from $50$ to $70$,
$d n_S/d \ln k$ varies from about $-0.00115$ to
almost $-0.00059$.

It is important to study the ratio of tensor to scalar
perturbations as a function of $n_S$ to check whether or not this
ratio lies within the observational sigma bounds.\cite{14}
From (\ref{88}), one can obtain in three-space dimension
\begin{equation}
\label{94}
n_S-1 = - \frac{3R}{16}+\frac{9}{2C}.
\end{equation}
To plot $n_S$ and $R$ together with the $2D$ posterior constraints
in the $n_S-R$ plane, we need to run the CAMB program coupled to
the CosmoMc (Cosmological Monte Carlo) code, for more details see
Ref.[14] and references therein. We will study this problem in our
future works.

\section{Conclusions}

In previous our studies about time variable spatial dimension
model\cite{1} we showed that based on observational bounds on the
present-day variation of Newton's constant, one would have to conclude
that the spatial dimension of the Universe when the Universe was at the
Planck length to be less than or equal to $3.09$. In [1] we concluded
that if the dimension of space when the Universe was at the Planck scale
is constrained to be fractional and very close to $3$, then the whole
edifice of TVSD model loses credibility.\cite{1}

In this paper, we have studied the effects of time variability
of the spatial dimension on the time evolution of inflaton field,
space dimension and scale factor for the potential $m^2\phi^2/2$.
We have also obtained general and explicit formulae for the scalar
spectral index and its running in TVSD or decrumpling model and then
applied them to two classes of examples of the inflaton potential.
The correction terms due to time variability of space dimension
depend on the e-folding number and the universal constant of TVSD
model which is $C$. The numerical calculations for the spectral index
and its running have been done for two classes of examples.

We have also discussed on dynamical solutions of inflaton field,
scale factor and space dimension in TVSD or decrumpling chaotic
inflation model. The outline of results has been
shown in Figs. 1, 2 and 3. Fig.1 compares inflaton field within the
framework of TVSD model and the constant three-space chaotic inflation.
It is seen time variability of space dimension causes that inflaton
field evolves slowly. This behavior of the scalar field can be seen by
Eq. (\ref{15}) where the effect of time variable dimension is like
decreasing friction term in this equation, or
subsequently causes that the scalar field changes slowly. In Fig. 2
the evolution of space dimension has been obtained. It is seen from
Eq. (\ref{11}) that temporal rate of space dimension is
proportional to the square of space dimension. So during the inflationary
epoch the Universe with $D_{P} = 10$ loses $6$ dimensions in
comparison to $D_{P} = 4$ which loses less than one. In other words,
$\Delta D_{\rm inflation}^{D_{P} = 10} = 6$ and $\Delta D_{\rm
inflation}^{D_{P} = 4} <1 $. From Fig.3 it is seen that dynamical
character of the space dimension increases the e-folding number.\cite{4}
Starting an Universe with higher dimensions at the Planck
epoch, it leaves inflationary phase later.\cite{3}

The final point that must be emphasized is about decrumpling model in
cosmology. The original motivation of this model presented in the pioneer
paper\cite{12} was based on an {\it ad hoc} assumption inspired from
polymer physics. It is quite possible that this part of decrumpling model
should be revised. However, just how this should be done is far from
obvious. The progress in decrumpling model can only be made if there
is a breakthrough in terms of finding a natural mechanism for varying
the spatial dimension in some alternative fashion to that which we
have considered.

\section*{Acknowledgments}
F. Nasseri thanks Amir and Shahrokh for useful helps.


\begin{thebibliography}{0}
\bibitem{1} F. Nasseri, ``{\it Limits on the Time Evolution of Space
Dimensions from Newton's Constant}'', {\tt hep-th/0412070}.
\bibitem{2} F. Nasseri and S.A. Alavi, ``{\it Noncommutative
Decrumpling Inflation and Running of the Spectral Index}'',
{\tt hep-th/0410259}, accepeted in Int. J. Mod. Phys. D.
\bibitem{3} F. Nasseri, Phys. Lett. {\bf B 538} (2002) 223,
{\tt gr-qc/0203032}.
\bibitem{4} F. Nasseri and S. Rahvar, Int. J. Mod. Phys. {\bf D 11}
(2002) 511, {\tt gr-qc/0008044}.
\bibitem{5} R. Mansouri and F. Nasseri, Phys. Rev. {\bf D 60} (1999)
123512, {\tt gr-qc/9902043}.
\bibitem{6} R. Mansouri, F. Nasseri and M. Khorrami, Phys. Lett.
{\bf A 259} (1999) 194.
\bibitem{7} C.L. Bennett, {\it et al.}, Astrophys. J. Suppl.
{\bf 148} (2003) 1, {\tt astro-ph/0302207}.
\bibitem{8} H.V. Peiris, {\it et al.}, Astrophys. J. Suppl. {\bf 148}
(2003) 213, {\tt astro-ph/0302225}.
\bibitem{9} D.J.H. Chung, G. Shiu and M. Trodden, Phys. Rev.
{\bf D 68} (2003) 063501, {\tt astro-ph/0305193}.
\bibitem{10} A. Kosowsky and M.S. Turner, Phys. Rev. {\bf D 52}
(1995) 1739, {\tt astro-ph/9504071}.
\bibitem{11} J.A. Peacock, {\it Cosmological Physics}, (Cambridge
University Press, Cambridge, 1999).
\bibitem{12} M. Khorrami, R. Mansouri and M. Mohazzab, Helv.
Phys. Acta {\bf 69} (1996) 237.
\bibitem{13} E.W. Kolb and M.S. Turner, {\it The Early Universe}
(Eddison-Wesley, New York, 1990).
\bibitem{14} S. Tsujikawa and A.R. Liddle, JCAP {\bf 0403} (2004)
001, {\tt astro-ph/0312162}.
\end{thebibliography}
\end{document}